\documentclass[aps,prl,twocolumn,superscriptaddress,nofootinbib,10pt]{revtex4-2}

\usepackage{graphicx} 
\usepackage{dcolumn} 
\usepackage{bm} 
\usepackage{slashed}
\usepackage{xcolor}
\usepackage{xspace}
\usepackage{amsmath}
\usepackage{amsfonts}
\usepackage{lmodern}
\usepackage{tabulary}
\usepackage{dsfont}
\usepackage{hyperref}
\usepackage{placeins}
\usepackage[letterpaper,top=1.0in, inner=0.5in, outer=0.5in,bottom=1.0in,,headheight=0pt,headsep=20pt,centering]{geometry}

\usepackage{lineno}

\usepackage{defs}

\begin{document}

\widetext
\leftline{MITHIG-MOD-25-002}
\leftline{Version 0}

\title{Unbinned measurement of thrust in \boldmath $\ee$ collisions at $\sqrt{s}$ = 91.2 GeV \\ with ALEPH archived data} 

\author{\href{https://ee-alliance.org/home/}{The Electron-Positron Alliance}: Anthony Badea} 
\email{badea@uchicago.edu}
\affiliation{Enrico Fermi Institute, University of Chicago, Chicago, IL 60637, USA}

\author{Austin Baty}
\affiliation{Department of Physics, University of Illinois Chicago, Chicago, IL 60607, USA}

\author{Hannah Bossi}
\affiliation{Laboratory for Nuclear Science, Massachusetts Institute of Technology, Cambridge, MA 02139, USA}

\author{Yu-Chen Chen}
\affiliation{Laboratory for Nuclear Science, Massachusetts Institute of Technology, Cambridge, MA 02139, USA}

\author{Yi Chen}
\affiliation{Department of Physics and Astronomy, Vanderbilt University, Nashville, TN 37235, USA}

\author{Jingyu Zhang}
\affiliation{Department of Physics and Astronomy, Vanderbilt University, Nashville, TN 37235, USA}

\author{Gian Michele Innocenti}
\affiliation{Laboratory for Nuclear Science, Massachusetts Institute of Technology, Cambridge, MA 02139, USA}

\author{Marcello Maggi}
\affiliation{Istituto Nazionale di Fisica Nucleare, Bari Division, BA 70126, Italy}

\author{Chris McGinn}
\affiliation{Laboratory for Nuclear Science, Massachusetts Institute of Technology, Cambridge, MA 02139, USA}

\author{Michael Peters}
\affiliation{Laboratory for Nuclear Science, Massachusetts Institute of Technology, Cambridge, MA 02139, USA}

\author{Tzu-An Sheng}
\affiliation{Laboratory for Nuclear Science, Massachusetts Institute of Technology, Cambridge, MA 02139, USA}

\author{Vinicius Mikuni}
\affiliation{National Energy Research Scientific Computing Center, Berkeley Lab, Berkeley 94720, USA}

\author{Matthew Avaylon}
\affiliation{Scientific Data Division, Lawrence Berkeley National Laboratory, Berkeley, CA 94720, USA}

\author{Patrick Komiske}
\affiliation{Laboratory for Nuclear Science, Massachusetts Institute of Technology, Cambridge, MA 02139, USA}

\author{Eric Metodiev}
\affiliation{Laboratory for Nuclear Science, Massachusetts Institute of Technology, Cambridge, MA 02139, USA}

\author{Jesse Thaler}
\affiliation{Laboratory for Nuclear Science, Massachusetts Institute of Technology, Cambridge, MA 02139, USA}

\author{Benjamin Nachman}
\email{nachman@stanford.edu}
\affiliation{Department of Particle Physics and Astrophysics, Stanford University, Stanford, CA 94305, USA}
\affiliation{Fundamental Physics Directorate, SLAC National Accelerator Laboratory, Menlo Park, CA 94025, USA}

\author{Yen-Jie Lee}
\email{yenjie@mit.edu}
\affiliation{Laboratory for Nuclear Science, Massachusetts Institute of Technology, Cambridge, MA 02139, USA}

\date{\today}%

\begin{abstract}
The strong coupling constant ($\alpha_{S}$) is a fundamental parameter of quantum chromodynamics (QCD), the theory of the strong force.  Some of the earliest precise constraints on $\alpha_{S}$ came from measurements of event shape observables, such as thrust ($T$), using hadronic $Z$ boson decays produced in $e^+e^-$ collisions.  However, recent work has revealed discrepancies between event-shape-based extractions of $\alpha_{S}$ and values determined using other experimental methods.  This work reexamines archived $e^+e^-$ data collected at a collision energy of $\sqrt{s}=91.2$ GeV by the ALEPH detector at the Large Electron-Positron Collider.  Modern machine learning techniques are used to correct for detector effects in an unbinned manner, allowing the $T$ distribution to be measured with higher granularity than previous ALEPH measurements.  The new measurement reveals a small but systematic shift towards larger values of $\tau=1-T$, and the potential implications of this shift for $\alpha_{S}$ extractions are illustrated by comparing to state-of-the-art theoretical calculations. In addition, the region of $-6<\log\tau<-2$, where poorly-understood non-perturbative effects are large, is compared to modern parton shower Monte Carlo simulations.  This measurement provides unique new inputs for $\alpha_{S}$ extractions and also improves constraints on phenomenological models of QCD dynamics such as parton fragmentation and hadronization.
\end{abstract}

\maketitle


The strong force is a key pillar of the Standard Model of particle physics and a necessary ingredient for many searches for physics beyond the Standard Model~\cite{Achenbach:2023pba}. Because of this, a precise understanding of the underlying theory of the strong force, quantum chromodynamics (QCD), is of upmost importance for particle physics experiments.  Although QCD describes a majority of the phenomena observed at hadron colliders like the LHC, many constraints on its behavior still rely on data from lepton colliders where no quarks or gluons are present in the initial state.  Previous \ee experiments at CERN and SLAC, which had collision energies greater than or equal to the $Z$ boson mass, allowed for a precise study of QCD across a wide range of energy scales.  This range of scales is particularly important for accessing both perturbative (P) and non-perturbative (NP) QCD effects. Although these experiments ran decades ago, the data they produced was archived to enable future studies of these collisions.  In this work, we seek to extract new insights from these legacy data by using modern analysis tools based on deep learning.

One of the most well-studied observables for high-energy \ee collisions is the event thrust ($T$)~\cite{PhysRevLett.39.1587}.  This observable characterizes the extent to which an event looks like a pair of recoiling quarks that subsequently form collimated sprays of particles called jets, versus a uniform spray of particles.  Recently, the thrust distribution has received renewed attention because of its sensitivity to a fundamental parameter of QCD, the strong coupling constant $\alpha_{S}$.  In particular, extractions of $\alpha_{S}$ based on the thrust distribution and similar event shape observables are currently in significant tension from the world-average $\alpha_{S}$ value~\cite{Benitez-Rathgeb:2024ylc, Benitez:2024nav, Abbate:2010xh, Nason:2023asn, OPAL:2011aa, ALEPH:2003obs, dEnterria:2022hzv, Huston:2023ofk, Zyla:2020zbs, ParticleDataGroup:2024cfk}. Thrust is also a valuable input for constraining parton shower Monte Carlo simulations (PSMCs) used in particle, nuclear, and astrophysics~\cite{Albrecht:2025kbb, Assi:2025ibi}.

A challenge with the interpretation of existing \ee thrust measurements is that they are binned, as a result of using binned matrix methods to correct for detector effects. A priori binning is suboptimal for unspecified downstream tasks. Different tasks, such as extracting $\alpha_{S}$ and tuning PSMC models may prefer different binning schemes. Furthermore, binned measurements use a binned detector response, which limits precision even in one dimension~\cite{Desai:2025mpy}. 
We address these challenges by performing an unbinned measurement of thrust using the OmniFold (OF) machine learning method~\cite{Andreassen:2019cjw,Andreassen:2021zzk}. 
The OF method iteratively removes detector distortions with classifiers that process continuous inputs to produce an unbinned, differential cross section estimate.  


In this work, OF is applied to archived \ee collision data collected at a center-of-mass energy of $\sqrt{s} = 91.2$ GeV by the ALEPH experiment at the Large Electron-Positron (LEP) Collider~\cite{Decamp:1990jra, Assmann:567226}.  Hadronic events are selected in the $Z\rightarrow q\bar{q}$ decay channel, as this channel provides the cleanest comparison to theoretical QCD calculations at the $Z$-pole. The data sample was collected in 1994 with a total integrated luminosity of 40~\ipb, corresponding to 1.36 million events. Only the 1994 dataset is utilized as corresponding archived Monte Carlo (MC) simulations of the detector are only available for that year. The archived data are converted to an MIT Open Data (MOD) format~\cite{Tripathee:2017ybi, Badea:2019vey, Chen:2021uws, Bossi:2025xsi}.  While unbinned measurements have been conducted for $ep$~\cite{H1:2021wkz, H1:2023fzk, H1:2024mox, H1prelim:2025}, $pp$~\cite{ATLAS:2024xxl, ATLAS:2025qtv, CMS-PAS-SMP-23-008, LHCb:2022rky, Song:2023sxb}, and $AA$~\cite{Pani:2024mgy} collisions, 
this is the first such measurement with \ee data. When partitioned to match the binning used by the ALEPH Collaboration for their measurement of thrust~\cite{ALEPH:2003obs}, the result shows a systematic broadening towards larger $\tau=1-T$ values. To investigate this shift, the measurement is re-binned for a high-granularity comparison to state-of-the-art theoretical calculations with varied $\alpha_{S}$ and NP scales. The comparison motivates  new theoretical $\alpha_{S}$ extractions including fits to the newly measured $T$ distribution. 

We also demonstrate the unique capability of the unbinned measurement to constrain both P and NP effects by re-binning the $\log\tau$ distribution in the peak region, corresponding to the cores of jets where the cross section is controlled by Sudakov logarithms of $\tau$. We show that the unbinned measurement supports a significantly finer partitioning of the distribution in the dijet peak region as compared to the previous ALEPH measurement~\cite{ALEPH:2003obs}.  The data provide discrimination power between different PSMC models of hadronization and parton showering. In the End Matter, comparisons to a more traditional unfolding method and a breakdown of the measurement uncertainty components are provided.


The ALEPH experiment was a multipurpose particle detector with a forward-backward symmetric cylindrical geometry and a near $4\pi$ coverage in solid angle~\cite{Decamp:1990jra, ALEPH:1994ayc}. It consisted of an inner tracking detector (ID) surrounded by a superconducting solenoid providing a 1.5 T axial magnetic field, electromagnetic and hadronic calorimeters, and outer muon detectors. The ID covered the range $20^\circ \leq \theta \leq 160^\circ$, where $\theta$ is the polar angle with respect to the beam. It consisted of a silicon strip vertex detector (VDET), multi-wire drift chambers (ITC), and a time projection chamber (TPC). A lead and proportional wire chamber based sampling calorimeter, positioned between the TPC and solenoid coil, provided electromagnetic (EM) energy measurements in the range  $13^\circ \leq \theta \leq 167^\circ$. The iron return yoke of the solenoid magnet was instrumented with plastic streamer tubes and iron slabs to provide hadronic calorimetry in the range $6^\circ \leq \theta \leq 174^\circ$. The muon detectors were located outside of the hadronic calorimeter and consisted of two double-layers of streamer tube chambers, providing coverage over almost the full solid angle and geometrically following the hadronic calorimeter footprint. The luminosity was measured using small-angle Bhabha scattering events and monitors near the ID and end-caps~\cite{Neugebauer:804776}. A two-level trigger system selected events from collisions. The first-level trigger used information from the calorimeters and ITC track candidates. The second-level trigger incorporated TPC track information to enhance event selection. The system was nearly fully efficient for collecting hadronic $Z$ boson events. An extensive software suite supported the data simulation, reconstruction, analysis, operations, trigger systems, and data acquisition. 

The information from the tracking detectors and the calorimeters are combined in an energy-flow algorithm~\cite{ALEPH:1994ayc}. For each event, the algorithm provides a set of charged and neutral reconstructed particles, called energy-flow objects, with measured momentum vectors and information on particle type. Good tracks are defined as charged particle tracks reconstructed with at least four hits in the TPC (NTPC $\geq 4$), originating from within a cylinder of length 20 cm and radius 2 cm coaxial with the beam and centered at the nominal collision point. The charged energy-flow objects are required to be contained within $|\cos\theta| < 0.94$ and have a minimum transverse momentum (\pT) of 0.2 GeV/$c$. The neutral energy-flow objects are required to be contained within $|\cos\theta| < 0.98$ and have a minimum energy of 0.4 GeV. A missing momentum vector is constructed to account for invisible objects, such as neutrinos, as $\Vec{p}_{\rm MET} = -\sum\nolimits_{\rm neu, chg} \Vec{p}$, where the sum over particle momentum $p$ includes both charged and neutral objects.

Hadronic events are selected by following the standard ALEPH choices~\cite{Barate:1996fi}, requiring at least 5 good tracks, at least 13 reconstructed charged or neutral particles, and a total charged energy $E_{\mathrm{ch}}$ 
in excess of 15~GeV. To ensure that an event is well contained within the detector, the polar angle of the sphericity axis is determined from all selected particles and required to satisfy $|\cos\theta_{\text{sph}}| < 0.82$.

Samples of MC simulations are used to correct the measurement for detector effects. Alongside the archived ALEPH data, a dedicated tune of $\textsc{Pythia}$ 6.1 MC is also archived~\cite{Sjostrand:2000wi}. The simulation includes both particle and detector-level events.
The events are generated at leading order (LO) in QCD and tuned with a dedicated ALEPH configuration. Electromagnetic (EM) initial state radiation (ISR) and final state radiation (FSR) are included in the generation. 
Modern MC simulations are used to evaluate the theoretical uncertainty from the choice of prior in the unfolding. Samples of $\textsc{Pythia}$ 8.230~\cite{bierlich2022comprehensiveguidephysicsusage}, $\textsc{Herwig}$ 7.1.5~\cite{B_hr_2008, bewick2024herwig73releasenote}, and $\textsc{Sherpa}$ 2.2.6~\cite{Bothmann2019} MC are produced with default settings and without EM ISR effects. The ALEPH detector simulation, however, is not currently available to create new detector-level MC samples.


Thrust is defined in the center-of-mass frame of an $e^+e^-$ collision as~\cite{PhysRevLett.39.1587}
\vspace{-1em}
\begin{align}\label{eq:thrust}
    T = \max_{\hat{n}} \frac{\sum_i |\vec{p}_i \cdot \hat{n}|}{\sum_i |\vec{p}_i|}\,,
\end{align}
\noindent where the sum is over all particles in the event and the maximum is over 3-vectors $\hat{n}$ of unit norm. The vector $\hat{n}$ that maximizes thrust is known as the thrust axis.
As the thrust distribution roughly follows a log normal distribution, $\mathrm{P}(1- T) \sim \exp\left[- C \log^2 (1-T)\right])$ for a constant $C$, a second variable is defined as $\tau = 1-T$. For detailed descriptions of the thrust distribution, see Refs.~\cite{Benitez:2024nav, Abbate:2010xh, Becher:2008cf}. Thrust is computed from all selected charged and neutral particles. 
We use an exact thrust calculation algorithm based on the $\textsc{Herwig++}$ implementation~\cite{Brandt:1978zm,B_hr_2008,BUCKLEY20132803}. Heuristic thrust calculations were used in the previous ALEPH measurement~\cite{ALEPH:2003obs}. No significant difference is observed between the exact and heuristic calculations, owing to the minimum requirement on the number of particles per event.


OF is used to correct the measurement for detector distortions, thereby unfolding from detector to particle-level~\cite{Andreassen:2019cjw}. OF is applied only to the single observable $\tau$. At the core of the method are two-stages of neural networks (NN) that are used to perform neural likelihood estimation and sequential reweighting. We use identical configurations for both networks. Technical details of the networks are provided in the End Matter. The unfolding is regularized by the number of OF iterations, analogous to the number of iterations in the D’Agostini method for Iterative Bayesian Unfolding (IBU)~\cite{dagostini2010improvediterativebayesianunfolding}, and the use of early stopping in the training. It has been shown that the number of iterations leads to increasing statistical uncertainty for IBU, while decreasing statistical uncertainty from bootstrapping for OF~\cite{falcão2025highdimensionalunfoldinglargebackgrounds}. Results are stable for 3–5 OF iterations; we use 5 iterations in the central values. As a consistency check, we apply the same unfolding used for the data to the detector-level MC sample and verify that the result reproduces the corresponding generator-level distribution within uncertainties.

The random initialization of NN weights introduces variation in the final measurement. To quantify this effect, we perform $N$ independent OF trainings with different random seeds and define this set of trainings as a single ensemble. We then generate $M$ such ensembles. For each ensemble, the median event weight across the $N$ trainings is taken as the unfolded result. Once a binning on a particular variable ($\tau$ or $\log\tau$ in this letter) is chosen, an uncertainty is measured as the standard deviation across the resulting $M$ binned distributions. We find that $N = 100$ and $M=10$
is sufficient to constrain the uncertainty to a subdominant level. For all systematic variations, the unfolding is performed using the same ensembling size with $N$ trainings per variation. This ensures that the impact of random network initialization is consistently accounted for.

Several corrections are applied implicitly within the unfolding procedure. First, the detector-level MC is derived from a subset of the particle-level MC after a hadronic event selection is applied. The first step of the unfolding process utilizes the detector-level MC. A unique matching exists between events at the detector-level and particle-level prior to the hadronic event selection. The weights obtained from Step 1 of OF are propagated to those corresponding particle-level events. Subsequently, the Step 2 of OF is performed using the full particle-level MC. This effectively unfolds the distribution to the particle-level phase space before the hadronic event selections. Second, the MC samples are generated with EM ISR, primarily for radiative return studies. However, the theoretical predictions we aim to compare with do not include EM ISR and optionally include FSR. Therefore, a cleaning procedure is applied to remove those contributions. The MC particle history is traced to identify and remove conversion electrons originating from radiated photons, which removes EM ISR and may remove some EM FSR. The cleaning is applied to both the particle and detector-level archived MC before the unfolding procedure. Third, the particle-level thrust is computed using all particles besides those removed from the cleaning. In doing so, any systematic shift in the thrust from tracking inefficiencies are captured in the difference between the quantity at particle and detector-level, which is then corrected within the unfolding.



The statistical uncertainty of the measurement is derived conservatively from two sources. First, the effect of finite data and MC sample sizes on the unfolding is evaluated with a bootstrap method~\cite{Efron1979Bootstrap}. For both data and MC, the OF ensembling and unfolding are repeated independently with bootstrap resampling. At the start of each repetition, every event is assigned an independent weight drawn from a Poisson distribution with unitary mean. The standard deviation across $10$ fully ensembled unfoldings provides the bootstrap uncertainty for data and MC separately. Given the relatively small statistical uncertainties, $10$ repetitions were found to be sufficient. Second, the statistical uncertainty on the final weighted events from OF is computed as the weighted Poisson error. 
Those uncertainties, and all subsequent ones, are computed for a given binning and added in quadrature.



Experimental systematic uncertainties are evaluated by varying the track and event selections in both data and MC. The thrust is recalculated for all variations that modify the particle level selections. We follow the standard ALEPH procedure for charged track and energy selections~\cite{Barate:1996fi, ALEPH:2003obs}. For charged tracks, the required NTPC is varied from 4 to 7 and the minimum \pT~from 0.2 to 0.4 GeV/$c$. With the nominal particle selections, the $E_{\mathrm{ch}}$ is varied from 15 to 10 GeV. We account for other experimental systematics differently than the standard ALEPH procedure. To account for potential mismodeling of neutral objects, the ALEPH Collaboration evaluated a systematic uncertainty by recalculating thrust without any neutral objects in the energy range $1<E<2$ GeV.  This variation effectively redefines the thrust observable to behave more like the charged-particle thrust, which is known to behave differently than the thrust calculated with all particles~\cite{Chang:2013iba}.  In this work, we attempt to evaluate this uncertainty without such a redefinition of the observable.  Motivated by our previous work studying the jet energy resolution in this dataset~\cite{Chen:2021uws}, we apply neutral particle energy scale (NES) and efficiency (NEE) variations.  These variations are only applied in MC to improve the modeling of neutral particle reconstruction effects, but they affect the final measurement after being propagated through the unfolding procedure. The NES is varied by $\pm 5\%$, which was found to cover the difference in the neutral particle energy spectrum observed between data and MC. The maximum deviation of these two variations from the nominal result is taken as the NES systematic uncertainty.  The effect of a mismodelled NEE is accounted for by randomly removing 2.5\% of the reconstructed neutral particles passing all selections in each event.  The removal probability was determined by examining differences in the neutral particle yields in data and MC. To account for the impact of the missing momentum, the thrust is recalculated including the $\vec{p}_{\mathrm{MET}}$. The unfolding procedure is repeated for each variation and the resulting change is taken as a systematic uncertainty. As mentioned, a systematic arises from the random initialization of the NN weights and is measured as the standard deviation across the binned distributions from the $M$ ensembles. 

\begin{figure}[t!] 
    \centering
    \includegraphics[width=0.48\textwidth]{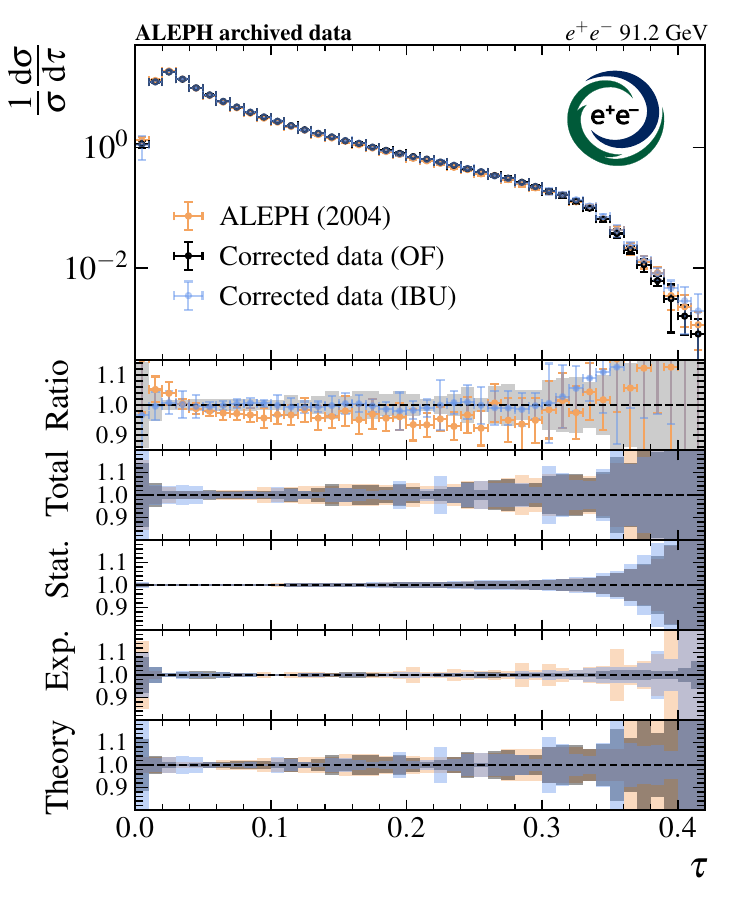}
    \caption{Unfolded thrust distribution with OmniFold (black) and Iterative Bayesian Unfolding (blue) using a binning consistent with the previous ALEPH publication (orange). The lower panels show the ratio to the OmniFold measurement and comparisons of the total, statistical, experimental, and theoretical uncertainties.}
    \label{fig:tau_multipanel}
\end{figure}

The theoretical uncertainty arising from the choice of MC prior used in the unfolding is measured by varying the prior and repeating the procedure. As the ALEPH detector simulation is not currently available to use for new simulations, a reweighting approach is used. The archived $\textsc{Pythia}$ 6.1 MC is reweighted at particle-level to the modern $\textsc{Pythia}$ 8, $\textsc{Herwig}$, and $\textsc{Sherpa}$ MC samples. The technical details of the reweighting are discussed in the End Matter. The full unfolding is repeated with the reweighting applied for each alternative MC. For each variation, the theoretical uncertainty is computed for a given binning choice as the difference from the nominal result. The final uncertainty is conservatively taken as the maximum bin-by-bin deviation across all variations. When evaluating the maximum difference, a quality check is applied: (1) bins from the theory variations are included only if the statistical uncertainty relative to the central value is $\leq 20\%$ and the pull with respect to the nominal result is less than or equal to 2 standard deviations, (2) if none of the theory variations pass the previous check for a particular bin then the minimum deviation is taken. These criteria ensure that the MC variations are reliable within the phase-space region relevant for the uncertainty estimate.

The unfolded thrust distribution is shown in Figure~\ref{fig:tau_multipanel} with a binning choice that matches the previous ALEPH publication~\cite{ALEPH:2003obs}. A complete breakdown of the uncertainties is shown in the End Matter, offering new insights into the components driving the total uncertainty beyond the grouped uncertainties reported previously by ALEPH. A full IBU measurement for the corresponding $\tau$ binning is also shown, with details given in the End Matter. The ratio between those measurements and the OF result is shown in the second panel. The total, statistical, experimental, theoretical MC prior uncertainties are shown in the subsequent panels. As mentioned, the uncertainty groups shown for the previous ALEPH publication represent the entirety of the available information on its uncertainty breakdown~\cite{hepdata.12794}. Our measured total uncertainty is less than 10\% for the bulk of the distribution and increasing in the tails. A shift is observed in the re-analysis towards larger values of $\tau$. The OF and IBU measurements agree, confirming that there is no systematic bias arising from the OF technique. Though this shift is interesting and may be physical, it is worth mentioning that there are notable differences in the dataset selection and analysis procedure between the measurements. We checked that the systematic shift is not removed at the data level by including the 1995 dataset or using the iterative thrust algorithm as implemented in the Rivet code from~\cite{hepdata.12794, Rivet:ALEPH_2004_I636645, rivet_thrust}.


\begin{figure}[t!]
\centering
\includegraphics[width=0.48\textwidth]{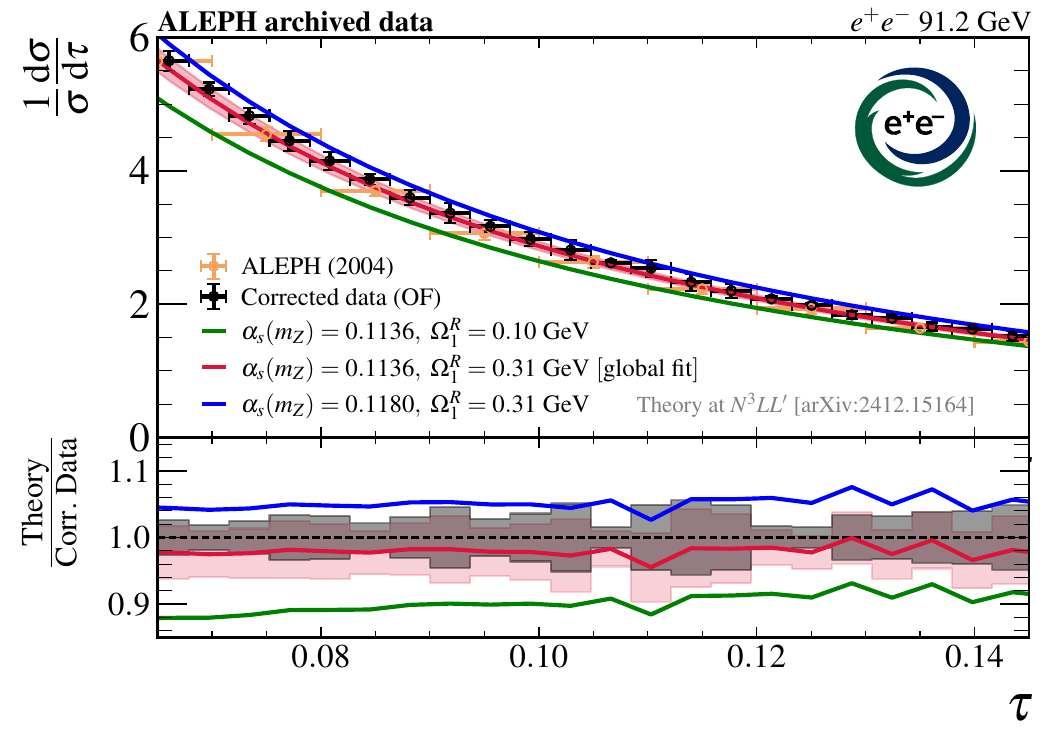}
\caption{Unfolded thrust distribution (black) in the tail region, compared with theoretical predictions varying $\alpha_{S}(m_{Z})$ and $\Omega_{1}^{R}$. The unbinned measurement is re-binned to the bin centers provided of the theoretical predictions. The global fit (red) shown reflects the best theory description of existing thrust data, excluding the measurement presented here.}

\label{fig:tau_tail_theoryComparison_WithRatio}
\end{figure}

The systematic shift toward larger $\tau$ values suggests that a dedicated theoretical fit to our measurement would yield a larger value of $\alpha_{S}$. The current discrepancy arises because theoretical fits to $e^{+}e^{-}$ event shapes predict a lower $\alpha_{S}$ than the world average~\cite{Benitez:2024nav}, meaning an upward shift could help resolve this tension. To test this, in Figure~\ref{fig:tau_tail_theoryComparison_WithRatio} we compare the measurement with state-of-the-art theoretical calculations directly in the $\tau$ tail region used for the theory fits. The authors of~\cite{Benitez:2024nav} provided theory curves on a fixed $\tau$ grid. The unbinned measurement is re-binned accordingly. The theoretical predictions are shown for varied values of $\alpha_{S}$ and $\Omega_{1}^{R}$, the first moment of a shape function modeling NP effects. The red curve and uncertainty band correspond to the global fit using existing data excluding our measurement, while the green and blue curves represent variations with $\Omega_{1}^{R}$ shifted down and $\alpha_{S}$ shifted up to the world average, respectively. The lower panel shows ratios of theory to our measurement, with uncertainty bands for both the data and the theory best-fit. Within uncertainties, the measurement agrees best with the global fit and shows some consistency with the upward $\alpha_{S}$ variation. However, since the theoretical calculation and fit procedure are highly nontrivial, no explicit conclusion about a definitive shift in $\alpha_{S}$ is drawn here. Instead, these results strongly motivate a dedicated theoretical study to perform a fit including our measurement.

\begin{figure}[t!]
\centering
\includegraphics[width=0.48\textwidth]{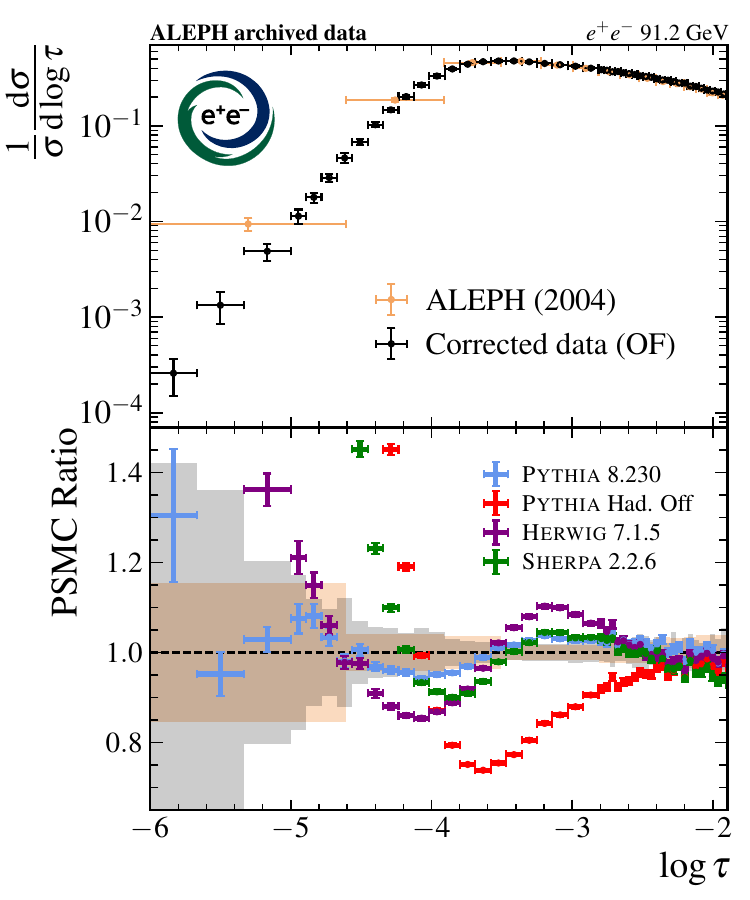}
\caption{Unfolded thrust distribution (black) in the dijet peak region in fine $\log\tau$ bins to highlight the sensitivity to non-perturbative effects in comparison to the logarithm of the previous ALEPH thrust publication (orange). Shown are comparisons to \textsc{Pythia} 8.230 with the Lund string hadronization model and $p_{\mathrm{T}}$ ordered dipole shower (gray), identical \textsc{Pythia} 8.230 except with hadronization disabled (red), \textsc{Herwig} 7.1.5 with the cluster hadronization model and angular-ordered shower (purple), and \textsc{Sherpa} 2.2.6 with the Lund string hadronization model (via \textsc{Pythia}) and the dipole shower (green).}
\label{fig:logtau_peak_nominal}
\end{figure}

The measurement is then re-binned in Figure~\ref{fig:logtau_peak_nominal} with fine $\log\tau$ bins to highlight the sensitivity to NP effects in the dijet peak region approaching $\tau\sim\Lambda_{\mathrm{QCD}}/\sqrt{s}$ or $\log\tau \sim -6$. The precision of the measurement degrades at lower values of $\log\tau$, limiting further investigations into the NP region $\tau<\Lambda_{\mathrm{QCD}}/\sqrt{s}$. The same figure with the OF measurement partitioned to the previous ALEPH publication binning is shown in the End Matter. The top panel, shows that the unbinned measurement supports much finer binning in the dijet peak region than the previous ALEPH measurement reported~\cite{ALEPH:2003obs}. As mentioned earlier, this region corresponds to the cores of jets where the cross section is dominated by Sudakov logarithms of $\tau$. The re-binning reveals the significant effects of the different hadronization and showering models in commonly used PSMC models, especially for $\log\tau < -2.5$. Shown are comparisons to \textsc{Pythia} 8.230 with the Lund string hadronization model and $p_{\mathrm{T}}$ ordered dipole shower (gray), identical \textsc{Pythia} 8.230 with hadronization disabled (red), \textsc{Herwig} 7.1.5 with the cluster hadronization model and angular ordered shower (purple), and \textsc{Sherpa} 2.2.6 with the Lund string hadronization model interfaced through \textsc{Pythia} and the dipole shower (green). We generally find that the measurement agrees better with the Lund string model of hadronization. The new fine binning, which provides discrimination power between the \textsc{Pythia} 8.230 models with hadronization on and off, motivates further experimental studies in this region to understand the true form of the NP shape function utilized in theoretical calculations.

In conclusion, the first unbinned thrust measurement in \ee collisions at a center-of-mass energy of $\sqrt{s}$ = 91.2 GeV using archived ALEPH data is presented. The thrust distribution is reconstructed using charged and neutral particles from hadronic $Z$ boson events. The measurement is compared with the previous ALEPH measurement in the corresponding binning in Figure~\ref{fig:tau_multipanel}, with the a systematic shift towards larger values of $\tau$ observed. This shift in $\tau$ could indicate a shift up in the value of $\alpha_{S}$, motivating future theoretical studies to assess if the effect can help resolve the existing discrepancies. The unbinned measurement is utilized to make high precision comparisons of the $\tau$ tail region to state-of-the-art theoretical calculations with varied values of $\alpha_{S}$ and $\Omega_{1}^{R}$ in Figure~\ref{fig:tau_tail_theoryComparison_WithRatio} and the $\log\tau$ peak region approaching $\tau\sim\Lambda_{\mathrm{QCD}}/\sqrt{s}$ to modern PSMC models with different hadronization and showering models in Figure~\ref{fig:logtau_peak_nominal}. The measurement provides unique constraints on both the P and NP regions. The full unbinned result in the form of event weights, thrust data, and analysis code are published alongside this letter at~\cite{ee_alliance_git, Badea:2025wzd}. This result paves the way for further precision studies of the strong force and the fundamental parameters of QCD with hadronic final states produced in high-energy \ee collisions~\cite{chen2024analysisnotetwoparticlecorrelation, Chen_2024, Benedikt:2651299, accardi2022opportunitiesprecisionqcdphysics}. 


\section*{Acknowledgments} The authors thank: Roberto Tenchini, Guenther Dissertori and Paolo Azzurri from the ALEPH Collaboration for their useful comments and suggestions on the use of ALEPH data, Iain Stewart and Miguel Benitez-Rathgeb for their insights and supplying the theory curves in Figure~\ref{fig:tau_tail_theoryComparison_WithRatio}, and Kyle Lee for discussions on the thrust resummation. This work has been supported by the Department of Energy, Office of Science, under Grant No. DE-SC0011088 (to Y.-C.C., Y.C., M.P., T.S., C.M., Y.-J.L.). A. Badea is supported by Schmidt Sciences. MA, VM, and BN are supported by the U.S. Department of Energy, Office of Science under contract numbers DE-AC02-05CH11231 and DE-AC02-76SF00515. This research used resources of the National Energy Research Scientific Computing Center, a DOE Office of Science User Facility supported by DOE contract number DE-AC02-05CH11231. This work was also supported in part by the National Science Foundation under Grant No. 2311666.

\bibliography{AlephLep1Thrust} 

\clearpage

\section{End Matter}
\label{sec:endmatter}

\begin{figure}[ht!]
    \centering
    \includegraphics[width=0.40\textwidth]{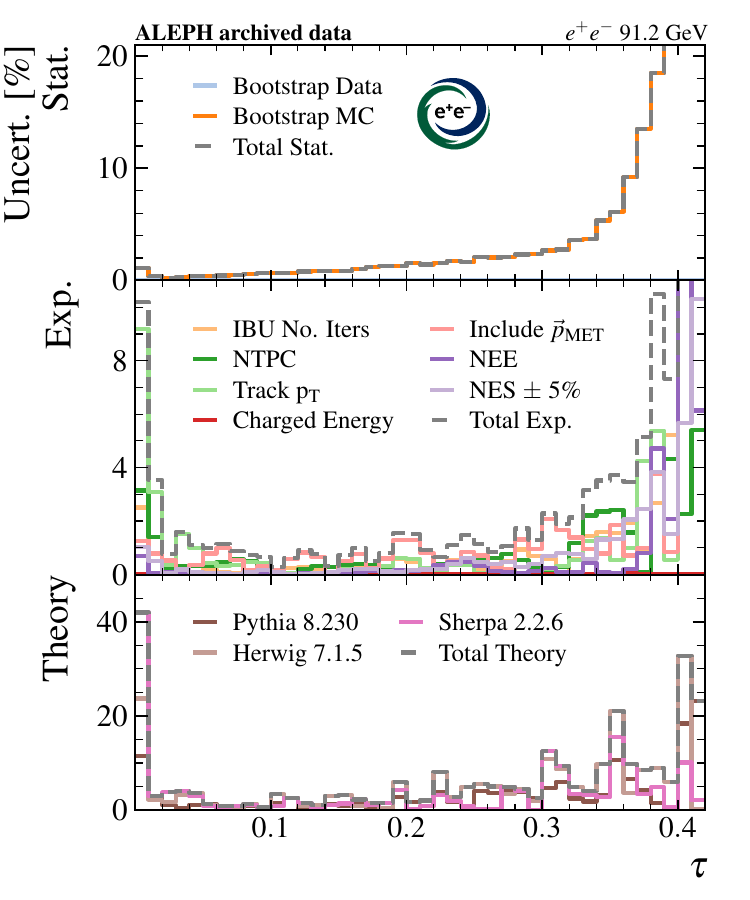} 
    \caption{Uncertainty breakdown for the fully corrected $\tau$ distribution, unfolded with Iterative Bayesian Unfolding in the previous ALEPH publication binning: (top) statistical, (middle) experimental, and (bottom) MC prior theoretical uncertainties.
    }
    \label{fig:IBU_tau_uncert}
\end{figure}

\begin{figure}[t!]
    \centering
    \includegraphics[width=0.40\textwidth]{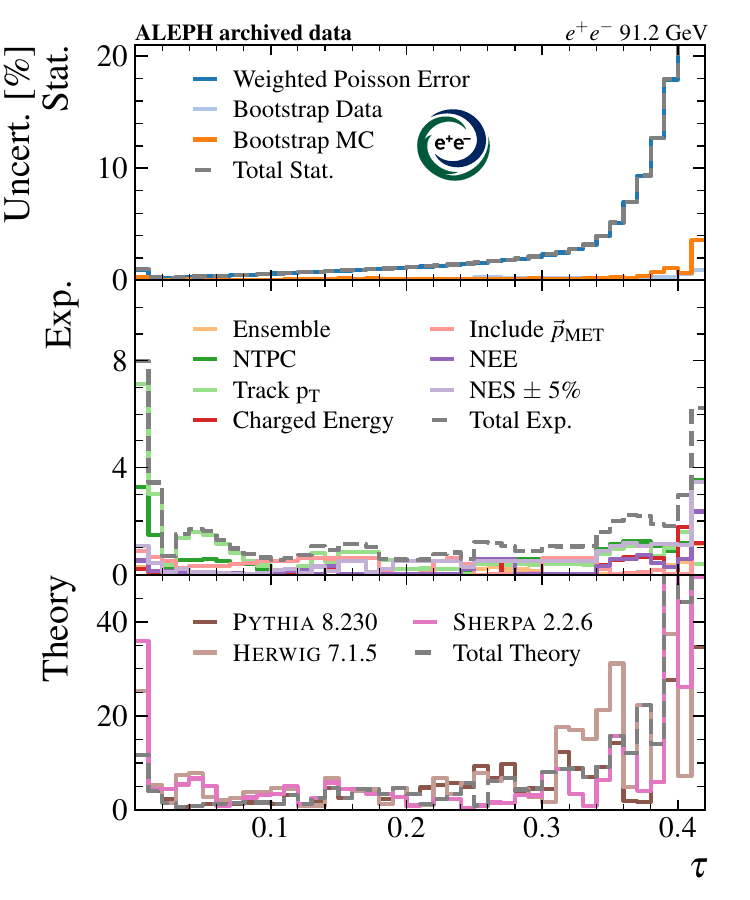} 
    \caption{Uncertainty breakdown for the fully corrected $\tau$ distribution, unfolded with OmniFold and partitioned in the previous ALEPH publication binning: (top) statistical, (middle) experimental, and (bottom) MC prior theoretical uncertainties.
    }
    \label{fig:OF_tau_uncert}
\end{figure}

\begin{figure}[t!]
    \centering
    \includegraphics[width=0.40\textwidth]{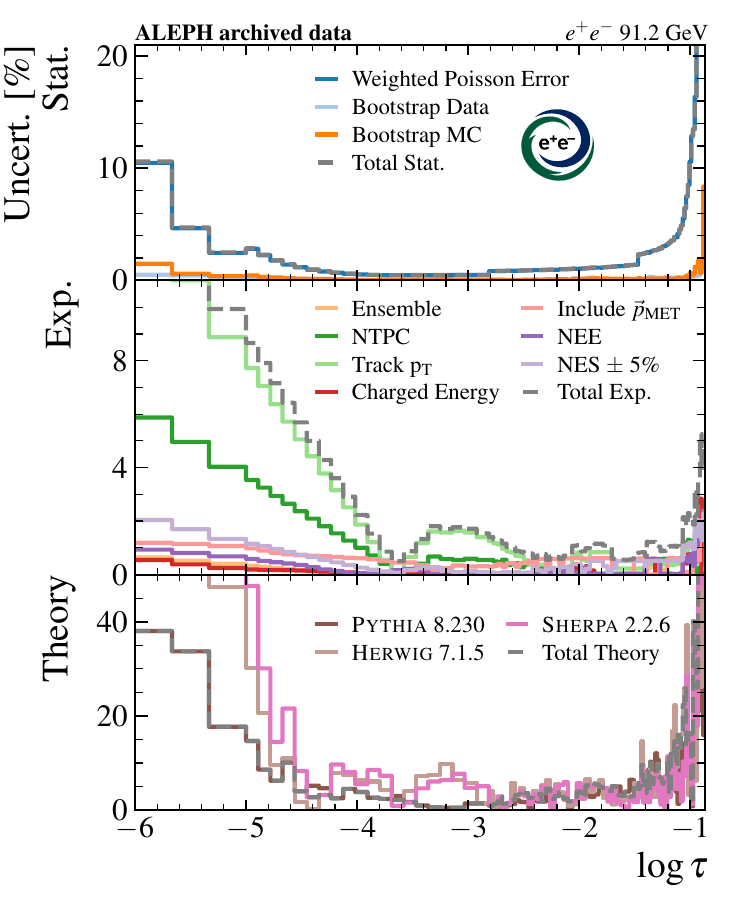} 
    \caption{Uncertainty breakdown for the fully corrected $\log\tau$ distribution, unfolded with OmniFold and partitioned in a fine binning: (top) statistical, (middle) experimental, and (bottom) MC prior theoretical uncertainties.
    }
    \label{fig:OF_logtau_uncert}
\end{figure}

\noindent \textit{\textbf{Uncertainty Breakdown:}} 
Figures~\ref{fig:IBU_tau_uncert} and~\ref{fig:OF_tau_uncert} show the OF and IBU measurement uncertainty breakdowns corresponding to the $\tau$ binning of Figure~\ref{fig:tau_multipanel}. Figure~\ref{fig:OF_logtau_uncert} shows the OF measurement uncertainty breakdown corresponding to $\log\tau$ binning of Figures~\ref{fig:logtau_peak_nominal}.  

\vspace{1em}

\noindent \textit{\textbf{OmniFold Training:}} The OF NN use identical configurations for both networks. The hyperparameters are optimized by fixing a subset of parameters and measuring the validation loss after the first OF step of the first iteration. The following settings are fixed to the algorithm’s default values based on prior usage: NNs with ReLU activations, training up to 100 epochs with early stopping triggered after 10 epochs if the validation loss plateaued. The width of the networks are scanned from 50 to 200 nodes with a fixed depth of 3 layers, while the depth is scanned from 2 to 4 layers with a fixed width of 100 nodes. Additionally, the batch size is scanned from $2^8$ to $2^{11}$, and the learning rate from $10^{-4}$ to $5\times10^{-3}$. The resulting validation losses are all comparable, leveling at around 0.602, indicating that the differences are minimal and insensitive to small changes. Based on these studies, we adopt the following standard configurations: networks with three dense layers with 100 nodes each, a batch size of 2048, and a learning rate of $5\times10^{-4}$. The fraction of the datasets that are used for training is 80\%, with the remaining data used for validation.
After training, the NN's are evaluated on the full datasets. All training is completed using TensorFlow software~\cite{tensorflow2015-whitepaper} running on NVIDIA A100 (NERSC Perlmutter) and RTX 4500 ADA Generation (UChicago DSI) GPUs.


\vspace{1em}

\noindent \textit{\textbf{Theory Reweighting:}}
In a similar manner to the steps of OF, the reweighting is learned at particle level via neural likelihood estimation using a Particle Edge Transformer (PET) classifier~\cite{Qu:2022mxj, shao2022transformerimplicitedgesparticlebased}. The PET is trained to distinguish events from the archived and alternative MC samples using the 
$(\log|p|, \eta, \phi)$ of all final state particles. First, the PET is pre-trained to distinguish the archived MC from itself, effectively initializing the model close to an identity mapping within the relevant phase space. Then, the primary training is carried out for up to 200 epochs with early stopping based on validation loss. The PET is then evaluated on the full particle-level MC sample. To mitigate the effects of random initialization, 15 PET's are trained and the predicted outputs are ensembled to yield the final reweightings. The full unfolding is then performed with the reweighting applied. The process is repeated for each of the alternative MC samples.

\vspace{1em}
\noindent \textit{\textbf{Iterative Bayesian Unfolding:}} We use IBU to unfold the thrust distribution in the binning of the previous ALEPH publication~\cite{ALEPH:2003obs}, allowing direct comparison to both the ALEPH measurement and the unbinned result partitioned to the same binning. A nominal choice of 2 IBU iterations is used. The result is stable for additional iterations but becomes more sensitive to statistical and binning effects. Systematic and theoretical uncertainties are evaluated similarly to the OF measurement by repeating the unfolding with varied spectra and comparing to the nominal result. An additional systematic uncertainty accounts for the impact of the number of iterations, taken as the difference between the nominal result and that from 3 IBU iterations. Statistical uncertainties are estimated via bootstrapping the Monte Carlo used to build the response matrix and the data, with the MC bootstrapping dominating the overall statistical uncertainty. Since the correspondence between generator- and detector-level archived MC exists only after the hadronic event selection, these samples are used to build the response matrices. To account for the selection, an explicit correction is applied after unfolding, computed from the ratio of the generator-level MC before and after hadronic event selection. This differs from the implicit correction approach used in the OF measurement. The same implicit correction for ISR as for OF is applied to the generator-level MC before the unfolding. No additional tracking inefficiency correction is applied.

\begin{figure}[ht!]
\centering
\includegraphics[width=0.40\textwidth]{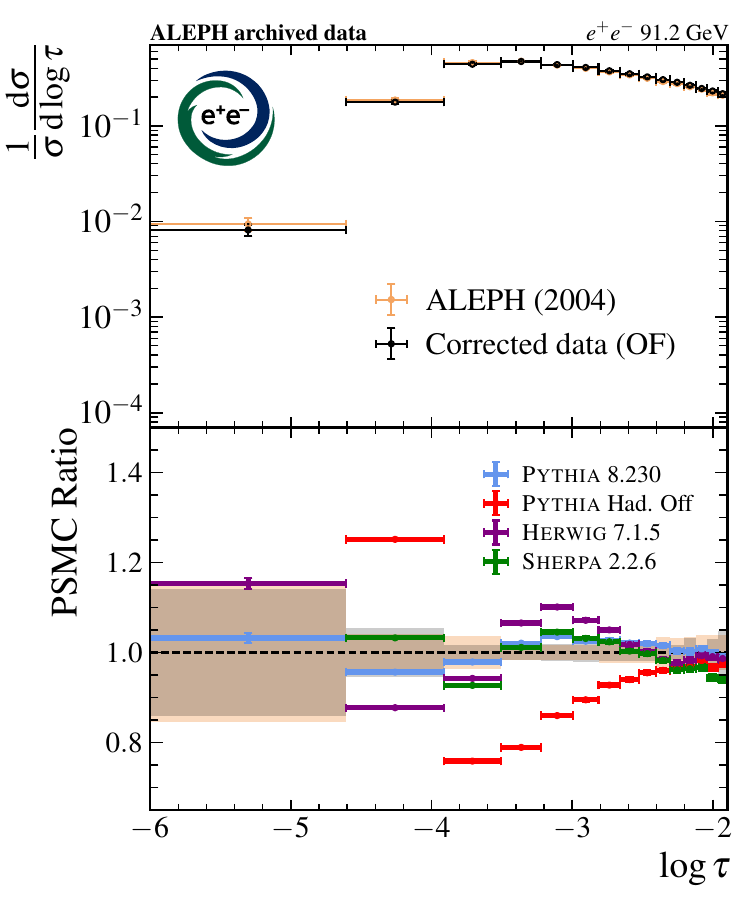}
\caption{Unfolded thrust distribution (black) in the dijet peak region partitioned in the bins corresponding to the previous ALEPH publication (orange). Shown are comparisons to \textsc{Pythia} 8.230 (gray), identical \textsc{Pythia} 8.230 except with hadronization disabled (red), \textsc{Herwig} 7.1.5 (purple), and \textsc{Sherpa} 2.2.6 (green).}
\label{fig:logtau_peak_aleph}
\end{figure}

\vspace{1em}
\noindent \textit{\textbf{Dijet Peak Region and NP Effects:}} Figure~\ref{fig:logtau_peak_aleph} shows the $\log\tau$ distribution for the OF measurement partitioned into the previous ALEPH publication binning. The ratio panel can be compared with the respective ratio panel of OF measurement partitioned into fine bins in Figure~\ref{fig:logtau_peak_nominal}. As shown in Figure ~\ref{fig:OF_logtau_uncert} as well, there is a trade-off between fine precision and granularity. Despite the increasing uncertainty, the fine granularity bins in Figure~\ref{fig:logtau_peak_nominal} reveals greater detail about the shapes of the NP effects approaching $\tau\sim\Lambda_{\mathrm{QCD}}/\sqrt{s}$ in comparison to the coarser bins of Figure~\ref{fig:logtau_peak_aleph}.

\clearpage

\end{document}